\documentclass [11pt]{article}
\usepackage{amsfonts,amsthm}
\usepackage{amsmath}
\newtheorem{theorem}{Theorem}
\newtheorem{proposition}{Proposition}
\newtheorem{lemma}{Lemma}
\newtheorem{corollary}{Corollary}
\date{}
\begin{document}
\title{Janossy Densities I. Determinantal Ensembles.
\footnote{ 
keywords and phrases: random matrices, orthogonal polynomials, Janossy densities, 
Riemann-Hilbert problem
}}
\author{ Alexei Borodin \thanks{Courant Institute of Mathematical Sciences, 251 Mercer Street, New York, NY 10012-1185.
Email address: borodin@caltech.edu.
This research was partially conducted during the period A.B.
served as a Clay Mathematics Institute Long-Term Prize Fellow.
}   
\and Alexander Soshnikov\thanks{
Department of Mathematics,
University of California at Davis, 
One Shields Ave., Davis, CA 95616, USA.
Email address: soshniko@math.ucdavis.edu.
Research was supported in part by the Sloan Research Fellowship and the 
NSF grant DMS-0103948. 
}  }
\date{}
\maketitle
\begin{abstract}
We derive an elementary formula for Janossy densities  for determinantal point 
processes with a finite rank projection-type kernel.
In particular, for $ \beta=2 $ polynomial ensembles 
of random matrices we show that
the Janossy densities on an interval $ I \subset \Bbb R$ 
can be expressed in terms of the  
Christoffel-Darboux
kernel for the orthogonal polynomials on the complement of $ I $.
\end{abstract}

\section{Introduction}

We consider an ensemble  of $ n $ particles  on a measure space $ (X, \mu)$
with the joint distribution density (with respect to the product measure $\mu^{\otimes n}$)
given by the formula
\begin{equation}
\label{Odin}
p(x_1, \ldots, x_n) = const_n \cdot \det(\phi_j(x_k))_{j,k=1,\ldots, n} 
\* \det( \psi_j(x_k))_{j,k=1, \ldots, n}\,.
\end{equation}
Here $ \phi_k(x), \  \psi_k(x), \ k=1, \ldots, n $, are some 
functions on $ X $
and  $ const_n $ is the normalization constant
\begin{eqnarray}
const_n^{-1} &=& \int_{X^n} \* \det(\phi_j(x_k))_{j,k =1, \ldots, n} 
\* \det(\psi_j(x_k))_{j,k=1, \ldots, n}
\* \prod_{j=1, \ldots, n} \mu(dx_j) \nonumber \\
\label{Dva}
&=& n!\, \det\left( \int_X \* \phi_i(x) \* \psi_j(x) \* \mu(dx) \right)_{i,j=1, \ldots, n}
\end{eqnarray}
where $ X^n = X \times \cdots \times X$ ($n$  times). Ensembles of this form were introduced in \cite{B}, \cite{TW2}. In the special cases when
$X=\Bbb R$, $\phi_i=\psi_i=x^{i-1}$, and $X=\{z\in \Bbb C\,|\,|z|=1\}$, $\phi_i=\overline{\psi}_i=z^{i-1}$, such ensembles were extensively studied in Random Matrix Theory much earlier under the general name {\it unitary ensembles}, see \cite{M} for details. An example of the form (\ref{Odin}) which is different from random matrix ensembles was considered in \cite{Mut}.

Let us assume that we can biorthogonalize
$ \{ \phi_j \}_{j=1, \ldots, n} $  and 
$ \{ \psi_j \}_{j=1, \ldots, n} $  with respect to the pairing
$$
\langle \phi,\psi\rangle=\int_X \* \phi(x) \* \psi(x) \* \mu(dx).
$$  
In other words, suppose that we can find functions
$ \xi_k(x), \ \eta_k(x), \ k=1, \ldots, n$ such that 
$$ 
\xi_k \in \textrm{Span}( \phi_j, \ j=1, \ldots, n), \ \ \eta_k \in \textrm{Span}( \psi_j, 
\ j=1, \ldots, n),\qquad  
\langle \xi_k,\eta_m\rangle = \delta_{km}.
$$
The families $ \{ \xi_j \} $ and 
$ \{ \eta_j \} $ are called
biorthogonal bases in
$ \textrm{Span}( \phi_j, \ j=1, \ldots, n) $ and  
$\textrm{Span}( \psi_j, \ j=1, \ldots, n) $ considered as subspaces in $ L^2(X, \mu).$
Then the distribution (\ref{Odin}) can be rewritten as (\cite{B}, \cite{TW2})
\begin{equation}
\label{plotnost}
p_n(x_1, \ldots, x_n)= \frac{1}{n!} \* \det(K(x_i, x_j))_{i,j=1, \ldots, n},
\end{equation}
with 
\begin{equation}
\label{Tri}
K(x,y)= \sum_{j=1}^n \* \xi_j(x) \* \eta_j(y).
\end{equation}
One of the particularly nice properties of the ensemble 
(\ref{Odin}), (\ref{plotnost})
is that one can explicitly calculate the correlation functions\\
$$ 
\rho_k(x_1, \ldots, x_k): = \frac {n!}{(n-k)!}\* \int_{X^{n-k}} \*
p(x_1,\ldots, x_k, x_{k+1},\ldots, x_n) \*
 \mu(dx_{k+1}) \cdots \mu(dx_n) 
$$ 
 which still have a determinantal form with the 
same  kernel $ K(x,y)$:
\begin{equation}
\label{det}
\rho_k(x_1, \ldots, x_k) =  \det (K(x_i,x_j))_{i,j=1,\ldots k}.
\end{equation}

If  $\mu$ is supported by a discrete set of points then 
the probabilistic meaning of the $k$-point correlation function is that of 
the probability to find a particle at each of 
$ k $ sites  $ x_1, x_2, \ldots x_k $. In other words,
$$ 
\rho_k(x_1, x_2, \ldots x_k) \mu(x_1) \cdots \mu(x_k)= 
\Pr \,\{ \textrm{there is a particle at each of the points}\ \  
x_i,\  i=1,\dots,k\}. 
$$
Analogously, if $X\subset \Bbb R$ and $\mu$ is absolutely continuous with 
respect to the Lebesgue measure then
$$ 
\rho_k(x_1, x_2, \ldots x_k) \mu(dx_1) \cdots \mu(dx_k)= 
\Pr \,\{ \textrm{there is a particle in each interval}\ \  
(x_i, x_i +dx_i)\}. 
$$

In general, random point processes with the $k$-point correlation functions of the determinantal form (\ref{det}) are called determinantal or fermion (see e.g. \cite{S1}).

So-called Janossy densities $\mathcal{J}_{k,I}(x_1, \ldots, x_k)$, $k=0,1,2, \ldots\, $, describe the distribution of the particles in a subset $I$ of $X$. If $X\subset \Bbb R$ and $\mu$ is absolutely continuous with respect to the Lebesgue measure then 
$$
\begin{aligned}
\mathcal{J}_{k,I}(x_1, \ldots x_k) \mu(dx_1) \cdots \mu(dx_k) =  
\Pr\,&\{\textrm{there are 
exactly $k$ particles in  $I$,}\\
{}&\textrm{ one in each of the $k$ infinitesimal 
intervals $(x_i, x_i +dx_i)$}\}.
\end{aligned}
$$

If $\mu$ is discrete then
$$
\begin{aligned}
\mathcal{J}_{k,I}(x_1, \ldots x_k) \mu(x_1) \cdots \mu(x_k) =  
\Pr\,&\{\textrm{there are 
exactly $k$ particles in $I$,}\\
{}&\textrm{ one at each of the $k$ points $x_i$}\}.
\end{aligned}
$$
See \cite{DVJ} for details.

For determinantal point processes  Janossy densities also have a determinantal form (see \cite{DVJ}, p.~140, or \cite{BO}, section 2):
\begin{equation}
\label{Janossy}
\mathcal{J}_{k,I}(x_1, \ldots, x_k)= const(I) \cdot 
\det(L_I(x_i,x_j))_{i,j=1,\ldots k},
\end{equation}
where 
\begin{equation}
 L_I= K_I(Id-K_I)^{-1}. 
\end{equation}
Here the kernel of $K_I$ is the restriction of the kernel $K(x,y)$ to $I$:
$K_I(x,y)= \chi_I(x) \* K(x,y) \* \chi_I(y)$, where 
$\chi_I(\,\cdot\,)$  is the characteristic function of $I$, and
$const(I)$ is the Fredholm determinant 
$$
const(I) = \det(Id-K_I)= \det (Id +L_I)^{-1}. 
$$ 

The main result of this paper is

\begin{theorem}
Let $ \widetilde\xi_j, \ j=1, \ldots, n $  and $\widetilde \eta_j, \ j=1, \ldots, n $ 
be biorthonormal bases in
$\textrm{Span} \{ \phi_j, \, j=1, \ldots, n \} $ and $\textrm{Span} \{ \psi_j, \, j=1, \ldots, n \} $ considered as subspaces of $ L^2(X \setminus I, \mu)$:
$$ 
\widetilde\xi_k \in \textrm{Span}( \phi_j, \, j=1, \ldots, n), \ \ \widetilde\eta_k \in \textrm{Span}( \psi_j, 
\, j=1, \ldots, n),\quad  
\int_{X\setminus I} \widetilde\xi_k(x)\widetilde\eta_m(x)\mu(dx) = \delta_{km}.
$$
Then the kernel of $L_I=K_I(Id-K_I)^{-1}$ is equal to
\begin{equation}
\label{kernelXI}
L_I(x,y)= \sum_{j=1}^{n} \widetilde\xi_j(x) 
\widetilde\eta_j(y).
\end{equation}
\end{theorem}

The above result readily applies to the so-called $\beta=2$ polynomial ensembles. Such ensembles arise, in particular, in random matrix theory \cite{M}, \cite{D1}, directed percolation and tiling models \cite{J1, J2, J3}, and representation theory \cite{BO}, \cite{BO-RSK}, \cite{BO-unit}. 
The definition is as follows.

Assume that $X$ is a subset of $\Bbb R$ and take
\begin{equation}
\label{polynom}
 p(x_1, \ldots, x_n)= const_n \cdot \prod_{1\leq j <k \leq n} (x_j-x_k)^2 .
\end{equation}
(Recall that this formula gives the joint distribution density with respect to $\mu(dx_1)\cdots\mu(dx_n)$.)

This is a special case of (\ref{Odin}) with 
$ \phi_j(x)=\psi_j(x)=x^{j-1}, \, j=1, \ldots, n$. Then we have
$\xi_j=\eta_j=p_{j-1}$, where $ \{p_j(x)\} $ are 
normalized orthogonal 
polynomials on $(X, \mu(dx))$, and $\deg(p_j)=j$.
The kernel $K(x,y) $ is the $n$th Christoffel-Darboux kernel 
\begin{equation}
\label{CD}
K_n(x,y)= \sum_{j=0}^{n-1} p_j(x) p_j(y) \nonumber 
         = \frac{k_{n-1}}{k_{n}} \* \frac{p_n(x)p_{n-1}(y)- 
p_n(y) p_{n-1}(x)}
{x-y}\,,
\end{equation}
where $k_j$ is the coefficient of $x^j$ in $p_j(x)$.
It should be noted that the kernel 
$K$ depends on $n$, but in what follows we will usually omit the subscript $n$ unless this may lead to a confusion.

Clearly, in the case of the polynomial ensemble (\ref{polynom}),  Theorem states that the kernel of $L_I=K_I(1-K_I)^{-1}$ is  the $n$th Christoffel-Darboux kernel computed for  the measure $\mu$ restricted to $X\setminus I$. That is,
$$
L_I(x,y)= \sum_{j=0}^{n-1} \widetilde p_j(x) \widetilde p_j(y) \nonumber 
         = \frac{\widetilde k_{n-1}}{\widetilde k_{n}} \* \frac{\widetilde p_n(x)\widetilde p_{n-1}(y)- 
\widetilde p_n(y) \widetilde p_{n-1}(x)}
{x-y}\,,
$$
where
$$
\widetilde p_j(x)=\widetilde k_jx^j+\{\textrm{lower order terms}\}, \qquad
\int_{X\setminus I}\widetilde p_k(x)\widetilde p_m(x)\mu(dx)=\delta_{km}.
$$

One of the particulary nice properties of the Janossy densities is that
for any interval (or, more genertally, a measurable set) $ I $
and non-negative integer $ k $ one has
$$
\Pr(\textrm{ there are exactly $k$ particles in $I$} )=  
\frac{1}{k!} \* \int_{I^k}  \mathcal{J}_{
k,I}
(x_1,\ldots, x_k) \mu(dx_1)
\cdots \mu(dx_k)
$$
The Janossy densities can be particularly useful in calculating the 
distribution of the left-most (right-most) particles when the particle space 
$X $ is a subset of the real line.  Indeed,
let us denote by $ \lambda_1 \leq \lambda_2\leq  \ldots \leq\lambda_n $ the 
locations of the particles in the increasing order. Then it is easy to see that
\begin{eqnarray}
\label{jankth}
& & \Pr(\lambda_{k} \in (s, s+ds))= \left( \frac{1}{(k-1)!} \*
\int_{(-\infty, s)^{k-1}}  \mathcal{J}_{
k,(-\infty, s)}
(x_1,\ldots, x_{k-1}, s) \mu(dx_1)
\cdots \mu(dx_{k-1}) \right)  \* \mu(ds)\nonumber\\
       &=&  \Pr( \lambda_1 \geq s) \* \frac{1}{(k-1)!} \*   
\int_{(-\infty,s)^{k-1}} \det(L_{(-\infty,s)}(x_i,x_j))_{i,j=1,\ldots k}\, 
\mu(dx_1) 
\cdots \mu(dx_{k-1}) \* \mu(ds)
 \\ 
\nonumber 
& & \Pr(\lambda_1 \geq s)= \left (\det(Id + L_{(-\infty,s)}) \right )^{-1}=
\det(Id -K_{(-\infty,s)}),
\end{eqnarray}
(where in (\ref{jankth}) we put $ x_k=s$).

This observation and the Theorem above allow us to compute explicitly the distribution functions of the left-most particles in the hard-edge scaling limit of random matrix models when the parameter (charge at the edge) is equal to zero. We refer to Section 4 below for the details. 

The result of Theorem 1 was initially discovered in the case of polynomial ensembles using the techniques of Riemann-Hilbert problems.
Later on, it was realized that Theorem 1 has a simpler linear algebraic proof. However, since the main idea of the ``Riemann-Hilbert'' computation is very useful in deriving Painlev\'e equations for the distribution of the left- or right-most particles in determinantal point processes, see \cite{BD}, \cite{B2}, \cite{BB}, we decided to include the argument into this paper; it can be found in Section 2. Section 3 contains the simpler proof. Concluding remarks are given in Section 5.

To conclude the Introduction, let us note that Theorem 1 has a counterpart for the so-called pfaffian ensembles. (The $\beta=1$ and 4 (or ``orthogonal'' and ``symplectic'') random matrix ensembles are the most known examples of the pfaffian ensembles.)
See the companion paper \cite{S2} for details. 

\section{Riemann-Hilbert Problem} In this section we will briefly describe two applications of the Riemann-Hilbert problem (to computing orthogonal polynomials and to inverting integrable integral operators) and use them to derive Theorem 1 in the case of polynomials ensembles. 
Since we use the Riemann-Hilbert problem (RHP, for short) mainly for instructional purposes, we avoid the discussion of any technical issues involved. 

Let $\Sigma$ be an oriented contour in $\Bbb C$.
We agree that when we go along the contour in the direction of orientation, the positive side lies to the left and the negative side lies to the right. 
Let $v$ be a map from $ \Sigma $
to $\textbf{GL}(l,\Bbb C)$, where $l=1,2,\dots$\,.
We say that an $l\times l$ matrix function $m=m(z)$ is a solution 
of the RHP $(\Sigma,v)$ if (\cite{CG}, 
\cite{D1})
\begin{eqnarray}
\label{R-HProblem}
 (i)&  &m(z) \ \ {\rm is} \ {\rm analytic} \ {\rm in} \ \  \Bbb C\setminus \Sigma, \\
(ii)&  &m_{+}(z)=m_{-}(z) \* v(z), \ \ z \in \Sigma.
\end{eqnarray}
Here $ m_{+}(z), \ m_{-}(z) $ stand for the limiting values of $ m(z) $ as  $z$ approaches $\Sigma$ from the positive (negative) side. If, in addition, $m(z) \to Id$ as $z \to \infty$ then we say that $m(z)$ solves the normalized RHP $(\Sigma,v)$. The matrix $v$ is usually called the jump matrix for the RHP.

First we describe the connection of RHP to orthogonal polynomials, see \cite{FIK1}, \cite{FIK2}.
Let $ d\mu(x)=\omega(x)\* dx $ be an 
absolutely continuous measure on the real line, such that the non-negative 
density 
$ \omega $ decays at infinity sufficiently fast (in particular, all moments 
exist).
Consider an RHP on $\Bbb R$ oriented from left to right with the jump matrix 
\begin{equation}
\label{jumpCD}
v(z)= \left ( \begin{array} {cc} 1 & \omega(z) \\ 0 & 1 
\end{array}
\right ).
\end{equation}
Fix a non-negative integer $n$. We are looking for a solution of the RHP $(\Bbb R,v)$ satisfying
$$ m(z)= \left ( Id + O(z^{-1})\right ) \* \left ( \begin{array} {cc} z^n & 0  \\ 
0 & z^{-n} \end{array} \right ),\qquad z\to\infty. 
$$

It appears that this RHP has a unique solution given by
\begin{equation}
\label{m}
m(z)= \left ( \begin{array} {cc} \pi_n(z) & (C (\omega \*\pi_n))(z) \\ 
\gamma_{n-1} \* \pi_{n-1}(z) & \gamma_{n-1} (C (\omega\pi_{n-1}))(z) 
\end{array} \right ), 
\ \quad z \not\in \Bbb R,
\end{equation}
where $ \pi_n(z)= z^n +\dots $ is the $n$th monic orthogonal polynomial corresponding to the weight function
$\omega(x)$, 
$$  
(C\*h)(z)= \frac{1}{2 \pi i} \int_{\Bbb R} \frac{h(\xi)}{\xi-z} \,d \xi 
$$ 
is the Cauchy transform, $\gamma_n= -2 \pi i \* k_n^2,$ and 
$ k_n$ is the leading coefficient
of the $n$th orthonormal polynomial $ p_n $, i.e. $ p_n(z)= k_n \* \pi_n(z)$. Thus, computing the orthogonal polynomials with the weight $\omega(z)$ is equivalent to solving RHP of the form above.

Now let us explain the relation of RHP to integrable operators.
Let $I$ be a subset of $\Bbb R$ (typically, a disjoint union of finitely many intervals).
We recall that an integral operator $M$ in $L^2(I,dx)$ with the kernel  $M(x,y)$ is called integrable
(\cite{IIKS1}, \cite{IIKS2}, \cite{D2}) if 
\begin{equation}
\label{integrable}
 M(x,y) = \frac{\sum_{i=1}^l f_i(x) g_i(y)}{x-y} 
\end{equation}
for some $l=2,3,\dots$ and some functions $f_i$, $g_i$ on $I$.
We assume that $\sum_{i=1}^l f_i(x) g_i(x)=0$ so that the kernel has no singularity on the diagonal.

In particular, the formula for the Christoffel-Darboux kernel in the case of polynomial ensembles discussed in \S1 means that the operators $K$ and $K_I$ can be viewed as integrable operators in $L^2(\Bbb R,dx)$ and $L^2(I,dx)$ with $l=2$, and we may take
\begin{gather}
\label{f}
f_1(x)=\frac1{2\pi i \,k_n}\,p_n(x)\sqrt{\omega(x)}, \quad f_2(x)=-k_{n-1}p_{n-1}(x)\sqrt{\omega(x)},\\ 
\label{g}
g_1(x)=2\pi i{k_{n-1}}p_{n-1}(x)\sqrt{\omega(x)},\quad
g_2(x)=\frac{1}{k_n}\,p_{n}(x)\sqrt{\omega(x)}.
\end{gather}

The appearance of $\sqrt{\omega(x)}$ has to do with the fact that we consider the Lebesgue measure rather than $\mu(dx)$ as our reference measure for the $L^2$-space.

It turns out that if the operator $Id - M$ is invertible then the resolvent $R=M \*(Id-M)^{-1}$ 
is also an integrable operator and 
\begin{gather}
\label{resolvent}
R(x,y)=\frac{\sum_{i=1}^l F_i(x) G_i(y)}{x-y}\,, \\
F_i   = (Id-M)^{-1} f_i, \qquad
G_i   = (Id-M^t)^{-1} g_i, \qquad \ i=1,2,\ldots, l,
\end{gather}
see 
\cite{IIKS1}, \cite{IIKS2} and also \cite{D2} 
for a very nice exposition. Furthermore, the functions $F_i$ and $G_i$ can be obtained through solving a RHP as follows. 
Let $v'$ be an $l \times l $ matrix valued function on $I$ 
given by
\begin{equation}
\label{jumpres}
v'= Id - 2 \pi i  \* fg^t,\qquad f=(f_1,\dots,f_l)^t,\quad
g=(g_1,\dots,g_l)^t.
\end{equation}
One can prove (\cite{IIKS1}, \cite{IIKS2}, \cite{D2}) that the normalized RHP $(I,v')$ has a unique solution $m'(z)$, and
\begin{eqnarray}
\label{ F,G}
F&=&(F_1,\ldots, F_l)^t =  (m')_{\pm} f\,,\\
G&=&(G_1,\ldots, G_l)^t = (m')^{-t}_{\pm} \,g.
\end{eqnarray}

The following observation is crucial.

\begin{lemma} Let $m$ be the solution (\ref{m}) of the RHP $(\Bbb R,v)$ and let $m'$ be the solution of the normalized RHP $(I,v')$ with $v'$ given by (\ref{jumpres}) and (\ref{f}), (\ref{g}). Then $M=m'm$ solves the RHP $(\Bbb R\setminus I,v)$ with the asymptotics $\textrm{diag}(z^n,z^{-n})$ as $z\to\infty$, and hence
\begin{equation}
\label{M}
M(z)= \left ( \begin{array} {cc} \widetilde\pi_n(z) & (C (\omega\chi_{\Bbb R\setminus I} \*\widetilde\pi_n))(z) \\ 
\widetilde\gamma_{n-1} \* \widetilde\pi_{n-1}(z) & \widetilde\gamma_{n-1} (C (\omega\chi_{\Bbb R\setminus I}\*\widetilde\pi_{n-1}))(z) 
\end{array} \right ), 
\ \quad z \not\in \Bbb R\setminus I,
\end{equation}
where $\ \widetilde{}\ $ signifies that the corresponding polynomials are orthogonal on $\Bbb R\setminus I$ with respect to the same weight function $\omega$.
\end{lemma}

The proof of this lemma is based on Lemma 4.3 of \cite{BD} (see also 
Lemma 2.4 of \cite{B2} for a discrete analog). The analog of Lemma 1 for weight functions with discrete support was one of the basic tools used in \cite{BB}. 
\begin{proof} A straightforward calculation shows that on $I$ we have $v'=m_+v^{-1}m_+^{-1}=m_-v^{-1}m_-^{-1}$. Thus on $I$
$$
M_-^{-1}M_+=m_-^{-1}(m'_-)^{-1}m'_+m_+=m_-^{-1}v'm_+=v^{-1}m_-^{-1}m_+=Id.
$$
On the other hand, since $m'(z)$ is holomorphic away from $I$ and tends to $Id$ as $z\to\infty$, it is clear that on $\Bbb R\setminus I$, $M(z)$ satisfies the same jump condition as $m(z)$, and that it also has the same asymptotics as $m(z)$ when $z\to\infty$. 
\end{proof}

\begin{proof}[Proof of Theorem 1] We apply the formalism described above to the Christoffel-Darboux kernel with $f_i$, $g_i$ specialized by (\ref{f}), (\ref{g}) above.
As before, we will use the notation 
$\widetilde p_k$, $\widetilde \pi_k$
for the $k$th orthonormal and monic orthogonal polynomials corresponding to the weight
$ \omega $ on $\Bbb R \setminus I$, and we also denote
$$
q_k=C (\omega\pi_k),\qquad \widetilde q_k=C (\omega \chi_{\Bbb R\setminus I}\widetilde\pi_k  ).
$$

In the calculations below we use the identity
$ \det m(z)\equiv\det M(z)\equiv 1$. Indeed, Liouville's theorem readily implies that if the jump matrix of a RHP has determinant 1 and the determinant of the asymptotics of a solution at infinity is also equal to 1, then the determinant of any solution of this RHP (having the corresponding asymptotics at infinity) must equal 1 identically.

We have
\begin{eqnarray*}
F_1&=& (m'f)_1=(Mm^{-1})_{11}f_1 + (Mm^{-1})_{12} f_2
      = M_{11}m_{22}f_1 - M_{12} m_{21} f_1
       - M_{11} m_{12} f_2 + M_{12}m_{11}f_2\\
      &=& \frac{k_{n-1}}{ k_n\widetilde k_n} \* \bigl ( -k_{n-1}
\widetilde p_n q_{n-1}
p_n  +\tilde{k}_{n}\widetilde q_np_{n-1}p_n \nonumber+ 
k_{n-1}\widetilde p_n q_n p_{n-1} - \tilde{k}_n\widetilde q_n p_np_{n-1} \bigr )\sqrt{\omega}
\nonumber\\
      &=&\frac{k^2_{n-1}}{k_n \widetilde k_n} \*( q_n p_{n-1}-q_{n-1} p_n) 
\widetilde p_n\sqrt{\omega} 
       \* 
       = \frac{1}{2\pi \* i\tilde{k}_n}  \,\det(m)\,\widetilde p_n\sqrt{\omega}
       =\frac{1}{2\pi \* i\tilde{k}_n}  \,\widetilde p_n\sqrt{\omega} . 
\end{eqnarray*}

Similar calculations yield
$$
G_1=2 \pi i  \* \widetilde k_{n-1}\widetilde p_{n-1}\sqrt{\omega},\quad
F_2=- \widetilde k_{n-1} \* \widetilde p_{n-1}\sqrt{\omega},\quad
G_2=\frac{1}{\widetilde k_n} \* \widetilde p_n\sqrt{\omega}.
$$
Hence, the kernel of $L_I=K_I(Id-K_I)^{-1}$ equals
$$
L_I(x,y)= \frac{F_1(x) G_1(y) + F_2(x) G_2(y)}{x-y} =
\frac{\widetilde k_{n-1}}{\widetilde k_n} 
\frac{\widetilde p_n (x) \widetilde p_{n-1}(y) - 
\widetilde p_n(y) \widetilde p_{n-1}(x)}{x-y}\,\sqrt{\omega(x)\omega(y)}.
$$
Recall that the factor $\sqrt{\omega(x)\omega(y)}$ is due to the fact that we are working in $L^2(I,dx)$ rather than $L^2(I,\omega(x)dx)$. The proof of Theorem 1 for polynomial ensembles is complete.

\end{proof}

\section{Linear algebraic proof}
We use the notation of \S1.
Consider the integral operators $ K_I $ and $ L_I $  in $L^2(I,\mu)$ with the kernels
\begin{equation}
\label{CD1}  
 K_n(x,y)= \sum_{j=1}^{n} \xi_j(x) \eta_j(y) 
\end{equation}
and 
\begin{equation}
\label{CDres1}
L_I(x,y)=  \sum_{j=1}^{n} \widetilde\xi_j(x) 
\widetilde\eta_j(y).
\end{equation}
Both operators  are finite-dimensional:
$$ 
\begin{gathered}
\textrm{Ran}( K_I) = \textrm{Ran} (L_I) = \mathcal{H}_1
:= \textrm{Span} (\phi_j)= \textrm{Span}(\xi_j)=\textrm{Span}(\widetilde\xi_j), \\
\textrm{Ker} (K_I) = \textrm{Ker} (L_I)= 
\mathcal{H}_2^{\bot},\qquad \mathcal{H}_2 := \textrm{Span}(\psi_j)=\textrm{Span}(\eta_j)= 
\textrm{Span}(\widetilde\eta_j), 
\end{gathered}
$$ 
where the index $j$ ranges over $\{1,\dots,n\}$, and the subspaces are taken inside $ L^2(I, \mu)$. 
Therefore, in order to prove that $ L_I = K_I (Id - K_I)^{-1}, $ it is enough 
to prove this relation for the restrictions of $ K_I $ and $ L_I $ to the 
$n$-dimensional space $ \mathcal{H}_1.$  To this end we 
compute the matrices of the restrictions of the operators $ K_I, \ L_I $
to $ \mathcal{H}_1 $ in the basis
$\{\xi_j\}_{j=1, \ldots n}$.

Let us denote by $ G_I $ and $G_{X\setminus I}$ the $ n \times n $
matrices with entries 
$$ 
(G_I)_{jk}=\int_I \* \xi_j(x) \* \eta_k(x) \* \mu(dx), \quad 
(G_{X\setminus I})_{jk}=\int_{X \setminus I} \* \xi_j(x) \* \eta_k(x) \* \mu(dx)
\qquad j,k=1, \ldots, n.
$$  
Since $ \{ \xi_j \}, \ \{\eta_k \} $ are biorthonormal on $X$,
we have $ G_I + G_{X \setminus I } = Id. $ The matrix of 
$ K_I $ on $ \mathcal{H}_1 $ in the basis $\{\xi_j\} $ 
is given by $ G_I.$  To calculate the matrix of 
the restriction of $ L_I $ on $ \mathcal{H}_1 $ we biorthonormalize the functions
$\xi_j$, $\eta_j$, $j=1,\ldots n$, in $ L^2(X \setminus I,\mu)$. This gives (cf. Proposition 2.2 in \cite{B})
\begin{equation}
\label{dokvo}
 \sum_{j=1}^{n} \widetilde\xi_j(x) \widetilde\eta_j(y) 
 = \sum_{j,k=1}^{n} \xi_j(x) \eta_k(y) 
(G_{X \setminus I})^{-1}_{kj}
 =\sum_{j,k=1}^{n} \xi_j(x) {\eta_k}(y) (Id -G_I)^{-1}_{kj}.
\end{equation}
It immediately follows from (\ref{dokvo}) that the matrix of the restriction 
of 
$ L_I $ to $ \mathcal{H}_1$ in the basis $\{\xi_j\}$ is equal  to $ G_I (Id -G_I)^{-1}$. The proof is complete.

\section{Hard edge with zero charge}

In a few special cases, the polynomials orthogonal with respect to
$ \omega $ on $ X \setminus I $ can be easily expressed in terms of the 
orthogonal 
polynomials on $ X $. We consider the Laguerre ensemble of positive 
definite  matrices as an example. 

Every positive definite $ n \times n $
matrix $M$ can be written (in a non-unique way) as $ M=AA^{*}$, where $ A $ is an
$ n \times n $ matrix with complex entries and $ A^*$ is the adjoint matrix. 
The probability measure in the
Laguerre ensemble (also called Wishart ensemble in statistics) 
is defined as (\cite{Br}):
\begin{equation}
\label{Wishart}
P(dM) = Z_n^{-1} \* \exp(- \textrm{Tr} (AA^*)) \det(AA^*)^{\alpha} \* dA,
\end{equation}
where $ dA $ is the Lebesgue measure on the $ 2 n^2$-dimensional space of
$ n \times n $ complex matrices, $ Z_n^{-1} $ is a normalization constant, and $ \alpha > -1.$
The joint probability density of the distribution of the eigenvalues of $M=AA^*$ is equal to
\begin{equation}
\label{Laguerre}
p(x_1, \ldots, x_n) = const_n \* \prod_{1 \leq i < j \leq n} (x_i-x_j)^2
\*  \prod_{j=1}^n x_j^{\alpha}e^{-x_j},
\quad \ x_j \in (0, +\infty), \ j=1, \ldots, n.
\end{equation}

The polynomials orthogonal with the weight $ \omega(x)=x^{\alpha}e^{-x}$
 on $\Bbb R_+=(0, +\infty) $ are the classical Laguerre polynomials (see e.g.
\cite{E}). In the special case of $ \alpha=0$ and $I=(0,t)$, the orthogonal 
polynomials on $ X \setminus I = \Bbb R_+  \setminus (0, t) = [t, +\infty) $ are
obtained from the Laguerre polynomials by the simple shift of variable $ x \mapsto x-t, $ i.e. $ \widetilde p_j(x)=
p_j(x-t)$, $j=0,1,\ldots$\,. Thus, by Theorem 1, the kernel of $L_I^{Lag(n)}=K_I^{Lag(n)}\left(Id-K_I^{Lag(n)}\right)^{-1}$ is equal to
\begin{equation}
\label{L}
L_I^{Lag(n)}(x,y)=K^{Lag(n)}(x-t,y-t),
\end{equation}
where $K^{Lag(n)}(x,y)$ is the order $n$ Christoffel-Darboux kernel for Laguerre polynomials with $\alpha=0$.

It is well known, see \cite{F}, \cite{NW}, \cite{TW1}, that when $n$ becomes large, the smallest eigenvalues in the Laguerre ensemble are of order $n^{-1}$. Moreover, if we rescale all the eigenvalues of the $n$th Laguerre ensemble by $n^{-1}$ then there exists a scaling limit as $n\to\infty$ of all the correlation functions. (In the random matrix theory this procedure is usually referred to as ``hard edge scaling limit''.) The limit correlation functions also have the determinantal form (\ref{det}) with the so-called Bessel kernel:
$$
\begin{gathered}
\lim_{n\to\infty} n^{-k}\rho_k^{Lag(\alpha,n)}\left(\frac {x_1}n,\dots,\frac{x_k}n\right)=
\det(K^{(\alpha)}(x_i,x_j))_{i,j=1,\dots,k}\,,\\
K^{(\alpha)}(x,y)=\frac{J_\alpha(2\sqrt{x})\sqrt{y}J_\alpha'(2\sqrt{y})-J_\alpha(2\sqrt{y})\sqrt{x}J_\alpha'(2\sqrt{x})}{x-y}\\=\int_0^1 J_{\alpha+1}(2\sqrt{\tau x})J_{\alpha+1}(2\sqrt{\tau y})d\tau
=\sum_{k,l=0}^\infty 
\frac{(-1)^kx^{k+\alpha/2}}{k!\Gamma(\alpha+k+1)}\,
\frac{(-1)^ly^{l+\alpha/2}}{l!\Gamma(\alpha+l+1)}\,
\frac{1}{\alpha+k+l+1}\,.
\end{gathered}
$$
Here $J_\nu(\,\cdot\,)$ is the J-Bessel function, see e.g. \cite{E}.
Note that if $\alpha=0$ then the above formula makes sense for any $x,y\in\Bbb C$.

\begin{proposition} For any $s>0$, let $K^{(0)}_s$ be the (bounded) integral operator in $L^2((0,s),dx)$ defined by the restriction of the Bessel kernel $K^{(\alpha)}(x,y)$ with $\alpha=0$ to $(0,s)\times(0,s)$. Then 
the operator $K^{(0)}_s\left(1-K^{(0)}_s\right)^{-1}$ is bounded and has a kernel which is equal to $K^{(0)}(x-s,y-s)$.
\end{proposition}
\begin{proof} The relation (\ref{L}) implies 
\begin{equation}
\label{L2}
K^{Lag(n)}(x,y)=K^{Lag(n)}(x-t,y-t)-\int_{0}^tK^{Lag(n)}(x-t,u-t)K^{Lag(n)}(u,y)du,\quad x,y\in(0,t).
\end{equation}
Since $n^{-1} K^{Lag(n)}(xn^{-1},yn^{-1})$ tends to $K^{(0)}(x,y)$ as $n\to\infty$ uniformly on compact subsets of $\Bbb C$ (this follows, e.g., from the proof of Theorem 4.5 in \cite{B}), taking the scaling limit in (\ref{L2}) yields
\begin{equation}
\label{L3}
K^{(0)}(x,y)=K^{(0)}(x-s,y-s)-\int_{0}^sK^{(0)}(x-s,u-s)K^{(0)}(u,y)du,\quad x,y\in(0,s).
\end{equation}
Denote by $L^{(0)}_s$ the operator in $L^2((0,s),dx)$ with the kernel $L^{(0)}_s(x,y)=K^{(0)}(x-s,y-s)$. Since the kernel of this operator is the uniform limit of the kernels of the nonnegative operators $L_I^{Lag(n)}=K^{Lag(n)}(x-t,y-t)$, we have $L^{(0)}_s\ge 0$, and hence $-1$ does not belong to the spectrum of $L^{(0)}_s$. Thus, (\ref{L3}) can be rewritten in the form $K^{(0)}_s=L^{(0)}_s\left(Id+L^{(0)}_s\right)^{-1}$, and this is equivalent to the statement of the proposition.
\end{proof}

\begin{corollary} Let $\lambda_1^{(n)}\le \lambda_2^{(n)}\le \dots \le \lambda_n^{(n)}$ be the ordered eigenvalues of the Laguerre ensemble (\ref{Laguerre}) with $\alpha=0$. Then 
\begin{gather}
\label{smallest}
\Pr\left(\lambda_1^{(n)}\geq \frac{s}{n}\right) = e^{-s},\\
\label{kth}
\lim_{n\to\infty}\Pr\left(\lambda_{k+1}^{(n)}\geq \frac{s}{n}\right)=e^{-s}\int_{(-s,0)^k}\det(K^{(0)}(x_i,x_j))_{i,j=1,\dots,k}\,dx_1\cdots dx_k,\qquad k\ge 2.
\end{gather}
In particular,
\begin{equation}
\label{second}
\lim_{n\to\infty}\Pr\left(\lambda_{2}^{(n)}\geq \frac{s}{n}\right)=\frac{e^{-s}}2 \int_0^{2\sqrt s} x(I_0^2(x)- I_1^2(x))dx,
\end{equation}
where $I_\nu(\,\cdot\,)$ is the I-Bessel function.
\end{corollary}

The formula (\ref{smallest}) was first observed in \cite{F}.
The limiting distribution (\ref{second})
of the second smallest eigenvalue was 
computed in \cite{TW1} and \cite{FH}. 
\begin{footnote}{Note that the integral (\ref{second}) can be evaluated 
in terms of Bessel functions, see e.g. (2.30) in \cite{TW1}.}\end{footnote} 
Further results in this direction, 
including formulas 
similar to (\ref{kth}) can be found in \cite{Wie}.

\begin{proof} The relation (\ref{smallest}) is easy:
$$
\begin{gathered}
\Pr\left(\lambda_1^{(n)}\geq \frac{s}{n}\right)=
\frac{\int_{(s/n,+\infty)^n} \prod\limits_{ i < j } (x_i-x_j)^2
\*  \prod\limits_j e^{-x_j}dx_j}{\int_{(0,+\infty)^n} \prod\limits_{i < j } (x_i-x_j)^2
\*  \prod\limits_j e^{-x_j}dx_j}=
\frac{\int_{(0,+\infty)^n} \prod\limits_{i < j } (x_i-x_j)^2
\*  \prod\limits_j e^{-x_j-s/n}dx_j}{\int_{(0,+\infty)^n} \prod\limits_{i < j} (x_i-x_j)^2
\*  \prod\limits_j e^{-x_j}dx_j}=e^{-s}.
\end{gathered}
$$
The relation (\ref{kth}) follows from (\ref{jankth}) applied to the Laguerre ensemble and the uniform convergence of kernels mentioned in the proof of the proposition above. 
Finally, using the L'H\^opital rule we obtain
$$
\int_0^s K^{(0)}(-x,-x)dx=\int_0^s \left(\left(J_0'(2i\sqrt{x})\right)^2-\frac {J_0(2i\sqrt{x})J_0'(2i\sqrt{x})}{2i\sqrt{x}}- J_0(2i\sqrt{x})J_0''(2i\sqrt{x})\right)dx.
$$
The formulas 
$$
J_0'(z)=-J_1(z),\quad J_0''(z)=z^{-1}J_1(z)-J_0(z),\quad I_0(z)=J_0(iz),\quad I_1(z)=-iJ_1(iz),
$$ 
see \cite{E}, and the change of variable $x\mapsto x^2/2$ bring the last integral to the form (\ref{second}).
\end{proof}
Of course, if $\lambda_1<\lambda_2<\dots$ are the ordered particles of the determinantal point process with the correlation functions given by the Bessel kernel with $\alpha=0$ then the right--hand sides of (\ref{smallest}) and (\ref{kth}) are equal to $\Pr\left(\lambda_{1}\geq {s}\right)$ and $\Pr\left(\lambda_{k+1}\geq {s}\right)$, respectively.

The calculations similar to those above can be done for the Jacobi ensemble corresponding to $ \omega(x)= (1-x)^{\alpha} (1+x)^{\beta}, \ x \in(-1,1)$, in the special 
cases $ I= (t,1)$, $\alpha=0$; $I=(-1,t)$, $\beta=0 $.
After appropriate rescaling one again obtains the limit relations of the form (\ref{smallest}),  
(\ref{kth}), (\ref{second}).

\section{Concluding remarks}
Take two $n$-point ensembles with joint probability densities of the form $const\cdot p_n(x_1,\dots,x_n)$ with the same $p_n$, but assume that these two ensembles are supported by different sets --- the first one lives on $(X,\mu)$ while the second one lives on $(X\setminus I,\mu)$, where $I$ is a subset of $X$. Of course, the normalization constants for these two ensembles will be different.
The $k$th Janossy density $\mathcal{J}_{k,I}(x_1,\ldots x_k)$ of the first ensemble is given by the formula
$$
\mathcal{J}_{k,I}(x_1,\ldots x_k)=const'\cdot\int_{(X\setminus I)^{n-k}}p_n
(x_1,\dots,x_n)\mu(dx_{k+1})\cdots \mu(dx_n),\qquad x_1,\dots,x_k\in I,
$$
while the $k$th correlation function $\widetilde \rho_k(x_1,\dots,x_k)$ of the second ensemble equals
$$
\widetilde \rho_k(x_1,\dots,x_k)=const''\cdot\int_{(X\setminus I)^{n-k}}
p_n(x_1,\dots,x_n)\mu(dx_{k+1})\cdots \mu(dx_n),\qquad x_1,\dots,x_k\in 
X\setminus I.
$$
The only difference between the two formulas above is in the constant prefactor and in the domain where $x_1,\dots,x_k$ are allowed to vary.
Clearly, this suggests that there should be a direct relation between $\mathcal{J}_{k,I}$ and $\rho_k$, and in the case of determinantal ensembles such relation is provided by Theorem 1.\begin{footnote}{Note, however, that a simple comparison of the two formulas above does not give a proof of Theorem 1 because these formulas only provide the symmetric minors of the corresponding kernels.}\end{footnote} 

Since the argument above does not depend on the specific form of the density $p_n$, one might expect that Theorem 1 should have an analog for the pfaffian ensembles (see e.g. \cite{R}, \cite{TW2}, 
\cite{Wid} for definitions). This is exactly the case, and the corresponding result is presented in the companion paper \cite{S2}.

\vspace{0.25cm}


\def\cmp{{\it Commun. Math. Phys.} }

\end{document}